# Long-lived nuclear coherences inside helium nanodroplets

B. Thaler[1], M. Meyer[1], P. Heim[1], and M. Koch[1*]

[1]Graz University of Technology, Institute of Experimental Physics, Petersgasse 16, 8010 Graz, Austria

*corresponding author: markus.koch@tugraz.at

**Much of our knowledge about dynamics and functionality of molecular systems has been achieved with femtosecond time-resolved spectroscopy. Despite extensive technical developments over the past decades, some classes of systems have eluded dynamical studies so far. Here, we demonstrate that superfluid helium nanodroplets, acting as thermal bath of 0.4 K temperature to stabilize weakly bound or reactive systems, are well suited for time-resolved studies of single molecules solvated in the droplet interior. By observing vibrational wavepacket motion of indium dimers ($In_2$) for over 50 ps, we demonstrate that the perturbation imposed by this quantum liquid can be lower by a factor of 10-100 compared to any other solvent, which uniquely allows to study processes depending on long nuclear coherence in a dissipative environment. Furthermore, tailor-made microsolvation environments inside droplets will enable to investigate the solvent influence on intramolecular dynamics in a wide tuning range from molecular isolation to strong molecule-solvent coupling.**

**Introduction.** A comprehensive understanding of mechanisms to convert solar energy into other energy forms is a prime objective for many research fields, with potential impact on light harvesting applications or the modelling of photoprotection in biomolecules. Despite extensive investigations, the primary processes triggered by photoexcitation of molecular systems remain insufficiently understood, as they proceed on pico- to sub-femtosecond time scales and involve concerted motion of electrons and nuclei in a complex manner. Insight into the functionality and dynamics of photoactive systems can be obtained in a unique way with femtosecond laser spectroscopy (*1*), revealing information about, for example, photofragmentation dynamics (*2*), molecular chirality (*3*), non-adiabatic coupling dynamics of electrons and nuclei (*4*), charge transfer (*5*), or electron dynamics (*6*). Recent experiments on chemically and biophysically relevant molecules suggest that nuclear motions and in particular their coherences have strong influence on the electronic evolution of the system (*7*); examples include prototypical molecules for photosynthesis (*8–10*) and photovoltaics (*11*).

The evolution of a molecular system after photoexcitation strongly depends on its immediate environment, with significant differences between isolated systems, where photodynamics can be precisely studied (*12, 13*), and the system in its real-world environment in condensed phase. Isolated molecules can be produced in a seeded supersonic expansion (*14*), where investigations are, however, often prevented by fragmentation resulting from excess energy during photoexcitation, or simply by the fact that weakly bound systems cannot be produced. This harmful vibrational energy can be dissipated to a thermal bath by embedding molecules in a high-pressure buffer gas (*15*) or a cryogenic matrix (*16*). As a disadvantage, influences of the environment on intrinsic dynamics can be severe and disentangling intra- and intermolecular dynamics is often impossible. Moreover, in such environments molecular dynamics cannot be probed with time-resolved photoelectron (PE) or -ion spectroscopy (*12, 13*), two very powerful methods that are independent of selection rules and dark states. Because of these drawbacks, many systems have eluded ultrafast studies.

Here we demonstrate that superfluid helium nanodroplets ($He_N$) are well suited to fill this gap as their influence on intrinsic dynamics of solvated molecules can be very low, and time-resolved PE detection can be used to observe dynamics of molecules located in the droplet interior. The benefits of $He_N$ for spectroscopy have been unveiled over the last three decades (*17, 18*): These nanometer-sized quantum fluid containers have enabled researchers to produce, isolate and investigate, for example, fragile agglomerates (*19, 20*), tailor-made complexes (*21, 22*), highly reactive species (*23*), or molecules in a controlled microsolvation environment (*24*). Femtosecond time-resolved investigations of $He_N$ have recently moved into the focus of researchers and various studies have been presented, including pure droplets (*25*), alkali-metal atoms and molecules located on the droplet surface (*26–28*), as well as alignment (*29*) and rotational studies (*30*). However, photoinduced dynamics of fully solvated molecules inside the droplet, where the majority of species is located, have not yet been studied in the time domain.

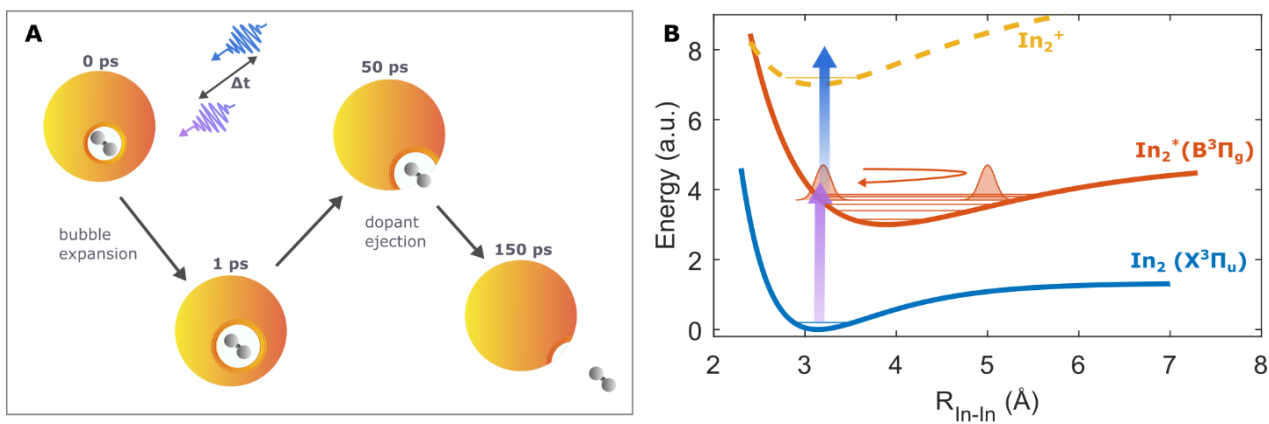

**Fig. 1. Schematic drawing of the photoinduced dynamics of the In$_2$-He$_N$ system.** (**A**) Solvent response: Expansion of the He solvation shell (bubble) during the first picosecond is followed by In$_2$ ejection within about 100 ps. (**B**) Intramolecular In$_2$ dynamics: A coherent superposition of vibrational states is generated by the spectrally broad pump pulse, leading to vibrational WP motion in the excited molecule. The WP is probed with a second pulse, resulting in a modulation of the photoelectron/ion yield due to an alternating ionization probability. Ground and excited state Morse potentials were taken from Ref. (*38*), the shape of the ionic potential is not known and therefore only schematically drawn.

Our results represent the first femtosecond investigation of molecular dynamics in the droplet interior, where the droplet influence is increased by orders of magnitude compared to surface-located species. To address the all-important question, namely the influence of the quantum fluid on intramolecular processes, we investigate vibrational WP dynamics of In$_2$ molecules. We disentangle these intramolecular dynamics from solvation dynamics by comparison to photoexcitation experiments of solvated atoms (*31*). Surprisingly, we observe that the decay of coherence in the In$_2$ nuclear WP is significantly reduced compared to conventional solvents.

**Results**. We generate single In$_2$ at 0.4 K temperature inside He$_N$ by sequential pickup of two In atoms. Photoexcitation with a pump pulse at the In$_2$ B$^3\Pi_g$ ← X$^3\Pi_u$ band (see Supplementary Note 1) triggers dynamics, which we trace by time-delayed photoionization with a probe pulse and PE/ion detection. We observe two different types of dynamics: The response of the He solvation shell (Fig. 1A) and a vibrational WP in In$_2$ (Fig. 1B). Both dynamics are represented in the transient PE spectra with overlapping time scales (Fig. 2) and we identify the solvation shell response by comparison with the In atom transient, which is described in our previous work (*31*).

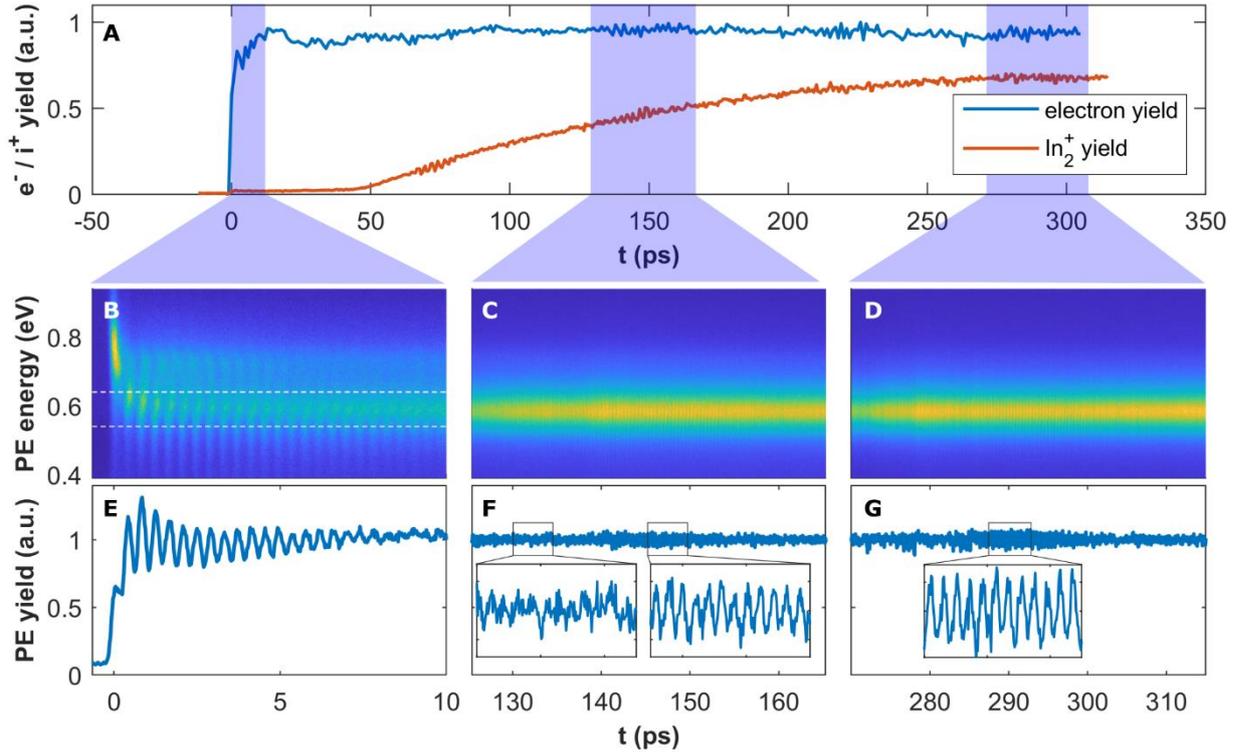

**Fig. 2. Photoelectron and -ion transients representing the dynamics of the In$_2$–He$_N$ system.** (**A**) Transient PE (blue) and In$_2^+$ (red) ion yields. (**B**) to (**D**) PE kinetic energy spectrum as function of pump-probe delay for temporal regions of the initial WP signal (**B**), as well as the first half (**C**) and full (**D**) revival, respectively; dashed lines mark the integration region. (**E**) to (**G**) Integrated PE signal for the energy region 0.54 to 0.64 eV, containing close ups of the half and full revival to better visualize the oscillating signals. Integrated curves are normalized to their sliding average (2.5 ps window), in order to compensate for long-term laser drifts.

Ground state atoms and molecules inside He$_N$ reside in cavities termed bubbles due to Pauli repulsion between the dopant's valence electrons and the closed shell He atoms. Photoexcitation and the correlated expansion of the valence electron orbital leads to an increase of the bubble size (Fig. 1A). For In$_2$, the connected transfer of potential energy to He kinetic energy is observed as shift of the PE peak from about 0.75 to 0.60 eV within the first picosecond (Fig. 2B). In the excited state, In$_2$ is ejected from the droplet, which can be deduced from the transient In$_2^+$ ion yield (red line in Fig. 2A). Following absent ion signals for the first 50 ps, the ion yield shows a slow rise within 200 ps because ionization of In$_2$ inside or in the vicinity of the droplet leads to trapping of the ion, preventing In$_2^+$ detection (*27*). We now turn to the intramolecular In$_2$ WP dynamics. Photoexcitation with a spectrally broad femtosecond laser pulse leads to coherent superposition of the vibrational eigenfunctions and the periodic movement of the resulting WP is detected as modulation of the PE signal (Fig.

1B). This modulation with a periodicity of 0.42 ps is clearly seen in the time-resolved PE spectrum (Fig. 2B), as well as the integrated PE yield (Fig. 2E). Anharmonicity of the potential leads to dispersion of the eigenfunctions and a spreading of the WP, detected as decrease of the modulation contrast (Fig. 2E). The reversible character of dispersion leads to refocusing of the WP at characteristic revival times, restoring the signal contrast to some extent. Half and full revivals of the $In_2$ motion are indeed observed around 145 ps (Figs. 2C & F) and 290 ps (Figs. 2D & G), respectively; the assignment stems from a comparison to the theoretical revival time, which is based on the anharmonicity parameter of the Morse potential (see Supplementary Note 1). The reduced amplitude in the revival signal, compared to the initial oscillation, reflects decoherence (dephasing) of the WP due to molecule-He interaction. Note that the full revival around 290 ps is observed at times where all excited dimers have left their droplets, as indicated by the leveling off of the $In_2^+$ ion signal (Fig. 2A). As expected, the next revival (3/2) can be observed around 435 ps, exhibiting the same oscillation contrast and temporal amplitude characteristics (not shown). The fact that fractional and full revivals are also observed in the $In_2^+$ yield (Fig. 2A), whereas they are absent during the initial oscillation, proves that the WP signals are associated to $In_2$ molecules that are originally solvated inside the droplet (see Supplementary Note 2).

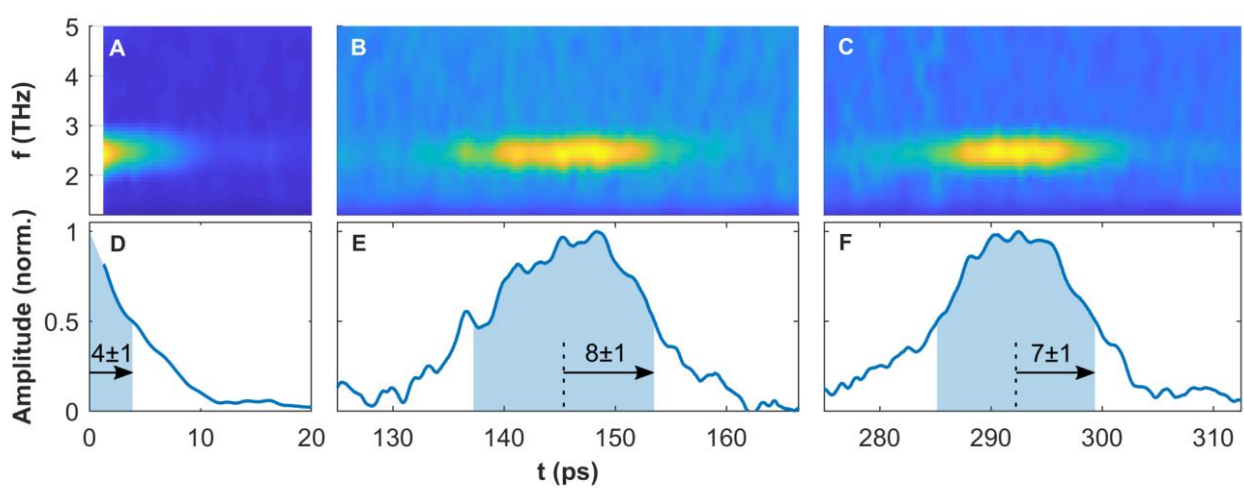

**Fig. 3. Sliding window fast Fourier analysis of the WP signals shown in Figs. 2B-D.** A Hamming window function of 2.5 ps width was used (see Supplementary Note 3), resulting in a spectral width of 0.55 THz (FWHM). (**A**) to (**C**): spectrograms, for which the PE energy range between 0.49 and 0.64 eV was considered. (**D**) to (**F**): time-dependent amplitudes of the central frequency (2.42 THz). Arrows indicate the time after which the signal has decreased to 50% of its maximum value (for the extrapolation to 0 ps in panel (**D**), see Supplementary Note 3).

Fourier transformation of the three datasets shown in Figs. 2B-D reveals always the same central frequency of (2.42 ± 0.05) THz, corresponding to an oscillation period of (0.41 ± 0.01) ps. To obtain insight into the transient changes of the oscillating signals, we apply sliding window Fourier analysis (see Supplementary Note 3). Figs. 3A-C show the three spectrograms corresponding to the initial WP oscillation, the half revival and the full revival, respectively. The transient amplitudes of the central frequency are shown in Figs. 3D-F, revealing that the signal amplitude of the initial WP oscillation (Fig. 3D) decreases faster than the amplitudes of the half and full revivals (Figs. 3E & F). The same frequency for $In_2$ inside and outside the $He_N$ hints at a minor influence of the helium solvent on the shape of the excited state potential. Upward-bent potentials, as observed for halogens in rare gas matrices (*16*), might however still influence WP motion at higher vibrational energies.

**Discussion**. We now further analyze the time- and frequency domain representations of the WP motion. The persistent strong oscillation signals within the first 10 ps (Figs. 2E & 3D) as well as the appearance of WP revivals (Figs. 2F & G and Figs. 3E & F) show that coherence is conserved to some degree inside $He_N$. Since the full revival occurs after the molecules are ejected from the droplet, comparison of the oscillation signal decay times for solvated and free molecules is possible and provides insight into the He-induced decoherence. Solvated $In_2$ experience both dispersion and He-induced decoherence of the WP, resulting in a 50% decrease of the oscillation amplitude within (4 ± 1) ps (Fig. 3D). Free $In_2$, in contrast, experience only dispersion, leading to a slower decrease of about (8 ± 1) ps (Figs. 3E & F). We thus estimate a decoherence half-life caused by He-interaction on the order of ~ 10 ps. This decoherence time is significantly shorter than the $In_2$ ejection time of ≥ 50 ps, pointing towards a non-constant, time-dependent loss of phase information; the vibrational WP dephases more strongly within the first few picoseconds after photoexcitation. Because the vibrational motion of $In_2$ is started at short internuclear distances (see Fig. 1B), one might expect a particularly strong energy transfer to the droplet during the first half oscillation period of 0.21 ps, as the atoms hit the solvation shell boundary. However, the photoexcitation-induced expansion of this shell takes also place within the first few hundred femtoseconds; the He boundary layer moves therefore at a similar pace as the separating In atoms, relativizing this assumption. A possible explanation for stronger initial He-influence might be found in helium density waves, which are initiated by the bubble expansion and reflected by the droplet surface to interact with the molecule after a few ps (*31*). Subsequent to this relatively strong initial interaction, decoherence during the molecule's propagation through the droplet seems to be much weaker. Even dopant ejection does not destroy the vibrational phase relations, despite possible recoil effects when the molecule ruptures the droplet boundary.

Viewing the In$_2$ and the He bubble as two quantum oscillators that are only weakly coupled because of their different excitation energies further suggests a low influence of the He surrounding: In$_2$ has a much larger excitation energy of 81 cm$^{-1}$ (0.41 ps oscillation period) compared to 1.1 cm$^{-1}$ (30 ps) of the He bubble (*31*). Coherence decay can be caused in two ways: first, pure dephasing of the vibrational modes (elastic dephasing), or second, vibrational relaxation within the In$_2$ excited electronic potential, resulting in energy transfer from the molecule to the droplet (inelastic dephasing). As phase-conserving vibrational relaxation would lead to a WP frequency increase (*32*), which is not observed in our case (Figs. 3A-C), we can conclude that for In$_2$ either solely elastic dephasing, or elastic dephasing in combination with non-coherent vibrational relaxation are present. It has been shown that vibrational relaxation rates depend on the energy mismatch between molecular vibrational modes and He droplet excitations (*21*). Recent simulations for I$_2$ excited to low vibrational states inside He$_N$ predict timescales down to a few hundred picoseconds (*33*), comparable to our measured In$_2$-He interaction times.

The low perturbing character of superfluid He as a quantum solvent becomes especially clear when the observed long-lasting vibrational coherences of ~50 ps are compared to other solvents, where coherence loss typically proceeds within some hundred femtoseconds up to a few picoseconds in special cases (*34*). Even in cryogenic rare-gas matrices decoherence times are limited to few picoseconds (*16*). For alkali metal dimers on He$_N$, which reside on the droplet's surface, a range of weak decoherence (~ 1.5 ns) (*35*) to strong decoherence (~ 5 ps) (*36*) was observed. As the dopant interaction with the He environment is increased in the droplet interior by two to three orders of magnitude compared to the surface, our finding of a low decoherence for fully solvated molecules is very surprising. The low influence on nuclear movement inside He$_N$ will be particularly important for systems with processes that are dominated by coherent nuclear motion (*7*), such as prototypical systems for photosynthesis (*8 –10*) and photovoltaics (*11*). Due to their confinement character, He$_N$ are able to isolate single donor-acceptor pairs of light-harvesting complexes (*37*). The generation of controlled microsolvation environments inside He$_N$ will allow to follow the transition from intrinsic dynamics of isolated molecules to the interaction-dominated behavior of solvated systems by successively adding solvent molecules (*24*).

In a general perspective, all dynamical studies in a dissipative environment face the problem that on the one side, coupling to a thermal bath is required to dissipate energy, while on the other side, transition states of chemical reactions are generally prone to increased solvent interaction as they are associated with large amplitude nuclear movements. Our results indicate that the strongly reduced influence of superfluid He might allow to follow transition state dynamics in many systems that were previously inaccessible. Coupling to the solvent depends on many aspects, such as the molecule's vibrational energy, the character of the vibrational mode, or the excited state electronic structure, influences that will be systematically characterized in future experiments.

## Materials and Methods.

We apply the helium nanodroplet ($He_N$) isolation technique, which has been explained in great detail elsewhere (*17, 18*). In short, $He_N$ are generated via continuous supersonic expansion of high purity (99.9999 %) helium gas through a cold nozzle. This expansion in combination with evaporative cooling leads to a temperature around 0.4 K, and therefore sample conditions well below the superfluid phase transition of $^4$He. For the presented experiments, the source conditions (5 µm nozzle diameter, 15 K nozzle temperature, 40 bar stagnation pressure) lead to a mean droplet size of about 9000 atoms. The droplets enter the pickup-chamber, where a resistively heated indium oven produces a vapor pressure leading to multi-atom pickup and the formation and stabilization of a dimer molecule inside a droplet. The doped droplets propagate to the measurement chamber, where they are investigated with femtosecond time-resolved pump-probe photoionization.

Laser pulses of 25 fs duration and 4.2 mJ pulse energy are obtained from a commercial amplified Ti:sapphire laser system (Coherent Vitara Oscillator and Legend Elite Amplifier) with 800 nm center wavelength at a repetition rate of 3 kHz. A one-photon pump, one-photon probe scheme with variable time delay is applied. Pump pulses are frequency up-converted via optical parametric amplification (Light Conversion, OPerA Solo) and subsequent frequency quadrupling to 345 nm (3.60 eV, 70 meV FWHM), in order to excite $In_2$ at the maximum of the in-droplet $In_2$ $B^3\Pi_g \leftarrow X^3\Pi_u$ transition band (see Supplementary Note 1). Probe pulses are frequency doubled to 406 nm (3.05 eV, 40 meV FWHM) in a 1 mm thick BBO crystal. Pump and probe powers are optimized to maximize the pump-probe signal contrast with respect to single pulse backgrounds, typical pulse energies lie in the range of a few µJ for both pulses. The cross correlation based on the $In_2$ overlap signal is estimated to be below 250 fs. Indium dimers are photoionized by the probe pulses and PE kinetic energies and ion charge-to-mass ratios are measured in a time-of-flight spectrometer. For photoelectrons the spectrometer is operated in a magnetic bottle configuration and for photoions a strong positive repeller voltage of about 2 kV is applied.

## Acknowledgements

We thank Wolfgang E. Ernst for providing the laser system and fruitful discussions, Leonhard Treiber for experimental support and Martin Schultze for reading the manuscript. We acknowledge financial support by the Austrian Science Fund (FWF) under Grant P29369-N36, as well as support from NAWI Graz.

## Supplementary Materials

**Supplementary Note 1: In$_2$-He$_N$ excitation spectrum and transition assignment.**

Fig. S1 shows excitation spectra of the indium monomer (In) and dimer (In$_2$) inside helium nanodroplets (He$_N$). The broad monomer excitation band around 27000 cm$^{-1}$ has been assigned to the atomic $6^2S_{1/2} \leftarrow 5^2P_{1/2}$ transition and is 2600 cm$^{-1}$ blue-shifted inside the droplets compared to the free atom (*39*). For In$_2$, we find two strong excitation bands in the investigated region. For the experiments shown within this work, excitation of the strong band around 29000 cm$^{-1}$ is applied, which we assign to an in-droplet $B^3\Pi_g \leftarrow X^3\Pi_u$ transition. A band for this transition, assigned by ab-initio calculations (*38*), has been measured in absorption around 27700 cm$^{-1}$ in cryogenic matrices (*40*), as well as in emission in a hot In gas (*41*). We conclude that the band exhibits a blue-shift of approximately 1300 cm$^{-1}$ (160 meV) inside the droplets, which is consistent with the measured transient photoelectron (PE) peak shift after photoexcitation (solvation bubble expansion within the first picosecond, see Fig. 2B in the main manuscript). The second prominent In$_2$ band centered around 27500 cm$^{-1}$ cannot be ascribed as easily. Its origin remains a speculation and might be due to the formation of a different dimer ground state inside the droplet (*38*) or to photoinduced dimerization of separated indium atoms after monomer excitation, similar to the behavior that has been described for magnesium (*42, 43*).

For the assigned $B^3\Pi_g$ excited state we now compare the measured wave packet (WP) parameters (oscillation period, dispersion time and revival time) to theoretical values derived from the fitted Morse-parameters of the ab-initio potential from Ref. (*38)*. We start with the general form of the Morse potential:

$$V(R) = D_e\left(1 - e^{-a(R-R_e)}\right)^2, \qquad (1)$$

where *D$_e$* is the dissociation energy, *a* the range parameter of the Morse potential and *R$_e$* the equilibrium bond distance. The fundamental frequency $\omega$ and the corresponding energy $\omega_e$ are defined by

$$\omega_e = \hbar\omega = \hbar a \sqrt{\frac{2D_e}{\mu}}, \qquad (2)$$

where μ is the reduced mass of the In$_2$ oscillator. The theoretical full revival time (*44*), meaning the time at which the vibrational eigenfunctions of the WP are again fully in phase, can be obtained with

$$T_{\text{rev}} = \frac{h}{\omega_e x_e}, \qquad (3)$$

where $\omega_e x_e = \omega^2 \hbar^2 / (4D_e)$ is termed the anharmonicity of the potential. As the degree of anharmonicity affects the initial dispersion of the wave packet, we can calculate the characteristic dispersion time $T_{\text{disp}}$ (meaning the time after which the wave-packet has completely dispersed; a derivation is presented in Ref. (*45*)) for a given central frequency *v* and the energetic spreading (FWHM) of the wave packet *ΔE*, by

$$T_{\text{disp}} = \frac{h^2 v(E)}{2\omega_e x_e \Delta E}. \qquad (4)$$

Table S1 compares the experimental time constants to the theoretical ones, which are calculated with the measured central frequency *v* = 2.42 THz (81 cm$^{-1}$) of the WP, the energetic bandwidth *ΔE* of the pump pulse (6.5 nm at 345 nm; corresponding to 550 cm$^{-1}$) and the Morse parameters of Ref. (*38*): $D_e$ = 1.14 eV and $\omega_e$ = 70 cm$^{-1}$.

**Table S1:** Comparison of theoretical and experimental values of the revival time (T$_{\text{rev}}$, Equ. 3) and the 50% value of the dispersion time (T$_{\text{disp}}$/2, Equ. 4). The experimental value for T$_{\text{disp/2}}$ is obtained as average of the half and full revival (see Fig. 3 in main manuscript).

|  | *T*$_{\text{rev}}$ (ps) | *T*$_{\text{disp/2}}$ (ps) |
|---|---|---|
| Theory | 250 | 9 |
| Experiment | 290 ± 5 | 8 ± 1 |

We describe the difference between theory and experiment for *T*$_{\text{rev}}$ to a deviation of the Morse model from the actual B$^3$Π$_g$ (II) potential. Nevertheless, the values are sufficiently similar for an assignment of the revivals: Based on the theoretical value of 250 ps, we assign the measured revivals at 145 ps and 290 ps to the half and full revival, respectively. Deviation of the Morse model is also reflected by the exceedance of the measured oscillation frequency of *v* = 2.42 THz (81 cm$^{-1}$) with respect to the predicted maximum value of *v* = 2.1 THz (70 cm$^{-1}$). The values for *T*$_{\text{disp}}$/2 (Tab. S1) are in agreement within the experimental uncertainty.

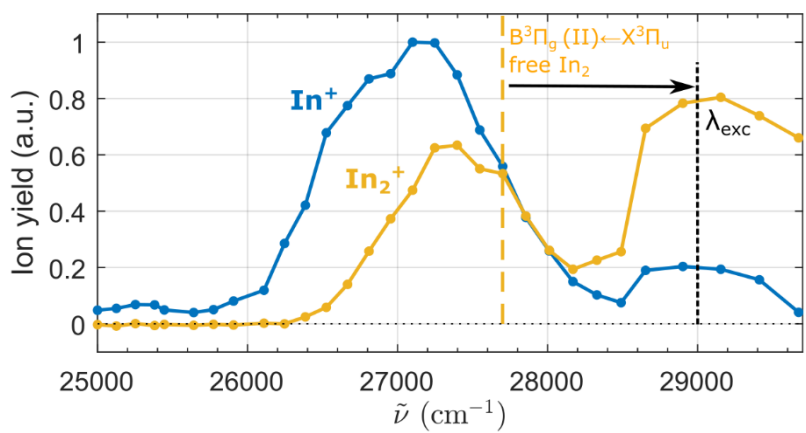

**Fig. S1: Excitation spectra of In and In$_2$ in helium nanodroplets.** The spectra are measured with pump-probe photoionization and ion detection, by using a fixed time-delay of 200 ps and scanning the wavelength of the pump pulse. The assigned transition of the free molecule is marked with the dashed yellow line. The black arrow indicates the blue-shift resulting from solvation inside the He$_N$ together with the excitation photon energy used for the experiments in this publication.

**Supplementary Note 2: Exclusion of a free In$_2$ background.**

The conclusions drawn from analysis of the In$_2$ WP are based on the assumption that all dimers are initially solvated inside He droplets and that no bare dimers are present in the droplet beam. These bare dimers would corrupt our results because they do not undergo He interaction after photoexcitation and are thus not subject to dephasing. Bare In$_2$ molecules could, in principle, originate from the pickup source, or be generated inside a He$_N$ and set free due to complete He evaporation in consequence of the cooling process. The most important argument that bare dimers do not contribute to the presented signals is that they cannot be photoexcited with the applied pump wavelength because the in-droplet excitation band is significantly blue shifted with respect to the bare In$_2$ excitation (see Supplementary Note 1). However, in the following we additionally present experimental verification of this assumption. Bare dimers originating directly from the pickup cell can be excluded because, first, the In vapor pressure inside the pickup source is far too low for molecule formation (< $10^{-3}$ mbar) and second, the PE signal vanishes when the valve between the He$_N$ source and the pickup region is closed. Free dimer production can be mediated by He$_N$, as both the double In pickup and In$_2$ formation dissipate significant amounts of energy into the droplet, potentially leading to complete He evaporation. This production mechanism can be analyzed by photoion yields at short time delays, as photoions produced inside $He_N$ cannot escape the strong trapping potential of the droplet, unless they have large kinetic energies. Figs. S2A & S2B show photoion mass spectra for the initial WP oscillations and the full revival, respectively, and Figs. S2C & S2D show transients of selected species. The $In_2$ mass signal at early times is very low and exhibits no oscillations, whereas at later times it dominates the mass spectrum with a pronounced oscillatory behavior. This strong signal increase confirms that the full revival oscillations are correlated to $In_2$ ejection from the droplet and that the contribution of initially bare $In_2$ is negligible.

Surprisingly however, we find that the early ion spectrum is dominated by $InHe_n^+$ (n = 0,..., ~ 30) clusters. We explain this ion signal with a fragmentation channel in the ionic state caused by the probe pulse. Ionization of the dimer partly proceeds to a dissociative state and substantial kinetic energy is released in the subsequent fragmentation process. At these early times ionization takes place inside the droplet and the produced $In^+$ can escape from the droplet due to their kinetic energy, while the attractive $In^+$-He interaction leads to the formation of $InHe_n^+$ clusters. As the fragmentation is probe-pulse induced, the $InHe_n^+$ yield is modulated with the neutral $In_2$ oscillation period. The very similar forms of the $InHe_n^+$ (Fig. S2C) and the total photoelectron transients (Fig. S2E) support this interpretation. The ionic fragmentation channel is also active for the free $In_2$ at long time delays, leading to oscillation signals in the $In^+$ yield, whereas $InHe_n^+$ complex formation is less likely (Fig. S2D). The overall ion yield at short delays is much smaller compared to that of the full revival, while the photoelectron signal is approximately the same, showing that only a fraction of the ion fragments leave the droplets. Interpretations in the main manuscript are based on transient photoelectron spectra and are thus not inuenced by this fragmentation process following the probe-ionization.

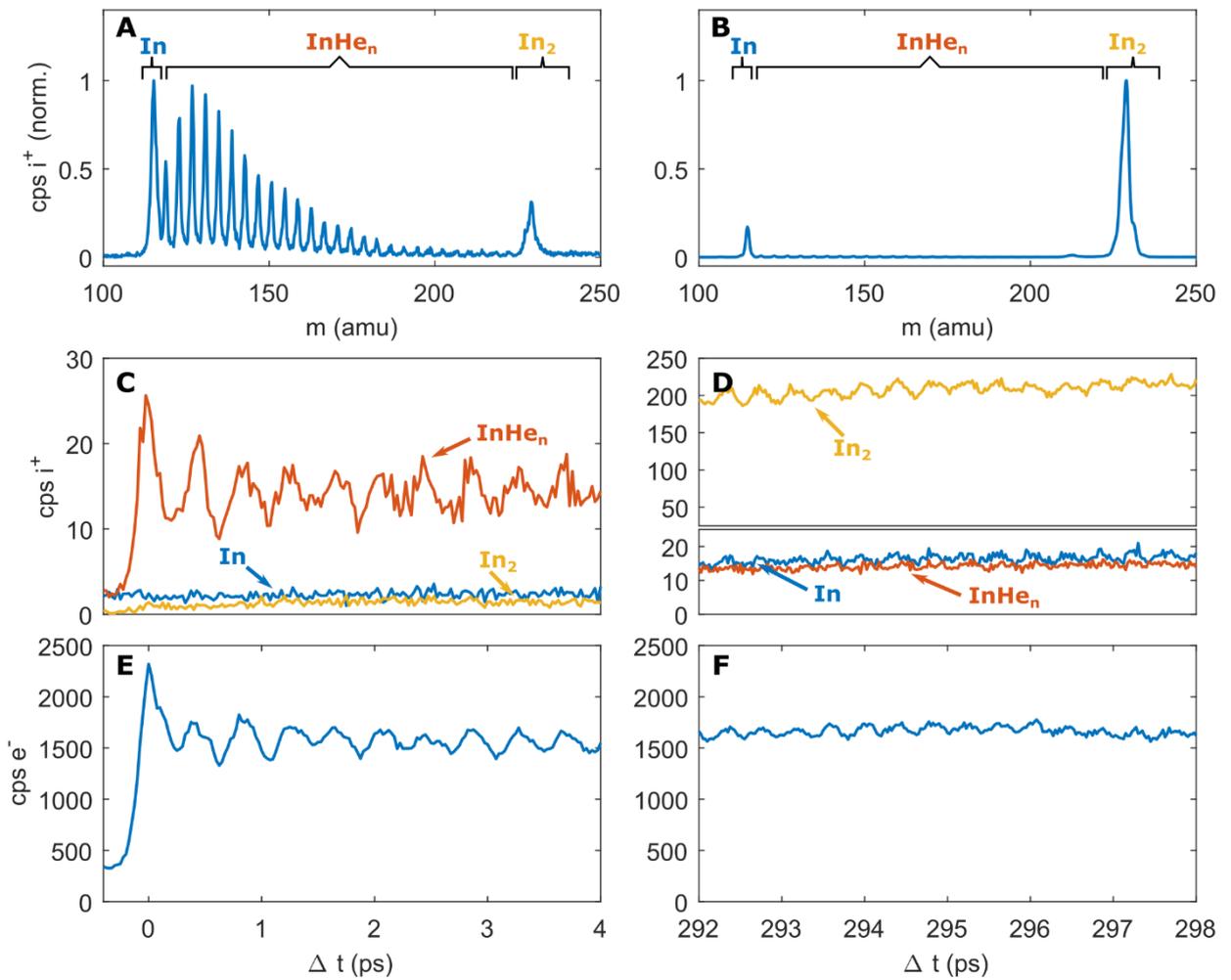

**Fig. S2: Comparison of photoion and photoelectron signals for the initial WP oscillation and for the full revival.** Panels (**A**) and (**B**) show mass spectra integrated within the time windows shown in (**C**) and (**D**), respectively. Panels (**C**) and (**D**) show the transient ion yields for mass intervals as indicated in (**A**) and (**B**). Panels (**E**) and (**F**) show corresponding time-resolved integrated electron yields.

**Supplementary Note 3: Sliding window Fourier analysis.**

The measured energy resolved photoelectron transients $S(t,E)$ (c.f., Fig. 2B-D) exhibit small oscillations on a big constant background. To increase the signal quality in the sliding Fourier analysis, we apply two adjustments to the transients: First, we analyze the spectrum only at time delays > 0 ps, as the steep edge around $t = 0$ ps makes it nearly impossible to detect small frequency components in Fourier space. Second, we apply a spectral filter at low frequencies in order to exclude problems arising from a strong peak at $f = 0$ THz, which stems from the constant offset in the PE transients and would blur the spectrum after convolution with the window function. Fourier transformation $\mathcal{F}\{S(t,E)\}$ of the whole measurement shows two distinct peaks, one at $f$ = 0 THz, as mentioned, and another at $f$ = 2.42 THz, which we assign to the WP motion. The transient PE spectrum is spectrally filtered with the function $F$ to obtain a filtered signal $S'$:

$$S'(t,E) = \mathcal{F}^{-1}\{\mathcal{F}\{S(t,E)\}\,F(f)\} \tag{5}$$

The filter function is defined with two Heaviside step functions Θ:

$$F(f) = \Theta(|f| - 1.4\ \text{THz}) - \Theta(|f| - 10\ \text{THz}) \tag{6}$$

The sliding window Fourier analysis is now performed by Fourier transformation of the filtered signal $S'(t, E)$ multiplied with a window function $W(t, t')$, resulting in an energy resolved spectrogram $\tilde{S}'(f, E, t')$:

$$\tilde{S}'(f,E,t') = \mathcal{F}\{S'(t,E)W(t,t')\} \tag{7}$$

A Hamming window function with a total width of $\Delta t$ is used to reduce spectral leakage:

$$W(t,t') = \left[\Theta\left(t - t' + \frac{\Delta t}{2}\right) - \Theta\left(t - t' - \frac{\Delta t}{2}\right)\right]\left[\frac{25}{46} + \frac{21}{46}\cos\left(\frac{2\pi(t-t')}{\Delta t}\right)\right] \tag{8}$$

Because the phase of the WP oscillation depends on the PE energy, we integrate only the amplitude of $\tilde{S}'(f, E, t')$ within the energy window between 0.49 and 0.64 eV to prevent interferences and finally obtain the desired spectrogram:

$$\tilde{S}'(f,t') = \int_{0.49\ \text{eV}}^{0.64\ \text{eV}} |\tilde{S}'(f,E,t')|\,dE \tag{9}$$

In the main manuscript, we extract the 50% decay time of the first WP signal by evaluation of the central frequency's (2.42 THz) transient amplitude. Because the abscissa resembles the center of the Hamming window, we linearly extrapolate the missing data points to 0 ps. Fig. S3 shows transient frequency amplitudes for different Hamming window sizes. The curve with Δt = 2.50 ps is fitted with a linear function from 1.25 to 3.2 ps, the resulting straight line extrapolates to 0 ps. Testing this approach for different window sizes and extrapolation regions results in the stated uncertainty (1 ps) of the 50% decay time, as is shown in Fig. 3 in the main manuscript. We note that in the main manuscript, the variable $t'$ of this derivation is termed $t$ for the sake of comparability.

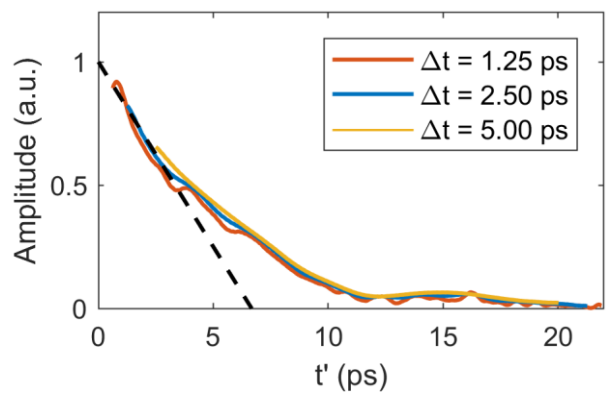

**Fig. S3. Transient WP amplitudes for different sliding window sizes.** Integrated Fourier amplitude of the *f* = 2.42 THz frequency with different Hamming window sizes *Δt* for the first wavepacket signals after photoexcitation. At short time delays the transient amplitude is linearly extrapolated.

**Supplementary References and Notes**